\documentclass[11pt]{article}
\usepackage[english]{babel}
\usepackage{setspace}
\usepackage{amsmath}
\usepackage{graphicx}
\usepackage[colorlinks=true, allcolors=blue]{hyperref}

\usepackage[utf8]{inputenc}
\usepackage[T1]{fontenc}
\usepackage[english]{babel}
\usepackage[a4paper,margin=2.5cm]{geometry}
\usepackage{graphicx}
\usepackage{amsmath,amssymb}
\usepackage[numbers]{natbib}
\usepackage{hyperref}
\usepackage{titlesec}
\usepackage{caption}
\usepackage{setspace}
\usepackage{multicol}
\usepackage{ragged2e} 
\usepackage[dvipsnames]{xcolor} 
\usepackage{tcolorbox} 
\usepackage{authblk}

\newcommand{\msun}{\,M$_{\odot}$}

\makeatletter

\renewcommand\AB@affilsepx{, }

\def\myinlineaffils{%
  {\small
   \renewcommand{\and}{, }
   \AB@affillist}
}

\renewcommand{\maketitle}{%
  \begin{center}
    {\LARGE\@title\par}
    \vskip 1em
    {\large
      \setlength{\parskip}{0.3em}
      \setlength{\parindent}{0pt}
      \@author
      \par}%
    \vskip 1em
    \vskip 1em
  \end{center}
}
\makeatother

\title{Why the Northern Hemisphere Needs a 30–40 m Telescope and the Science at Stake: \\ Mapping formation pathways of nuclear star clusters across galaxies} 

\author[1,2]{Francesca Pinna}
\author[3,4]{Isabel Pérez}
\author[1,2]{Anna Ferré-Mateu}
\author[1,2]{Begoña García Lorenzo}
\author[5]{Alessandra Mastrobuono Battisti}
\author[6]{Abbas Askar}
\author[1,2]{Michael Beasley}
\author[3]{Bahar Bidaran}
\author[7]{Ana L. Chies-Santos}
\author[2,1]{Sébastien Comerón}
\author[8]{Kristen C. Dage}
\author[2,1]{Adriana de Lorenzo-Cáceres}
\author[9]{Katja Fahrion}
\author[1,2]{Jesús Falcón Barroso}
\author[10]{Anja Feldmeier-Krause}
\author[1,2]{Emma Fern\'andez Alvar}
\author[10]{Nils Hoyer}
\author[11]{Rubén García Benito}
\author[11]{Rosa M. Gonzalez Delgado}
\author[1,2]{Ignacio Martín Navarro}
\author[1,2]{Cristina Ramos Almeida}
\author[12]{Patricia Sánchez Blázquez}
\author[13,1]{Rubén Sánchez Janssen}
\author[1,2]{Alexandre Vazdekis}

\affil[1]{Instituto de Astrofísica de Canarias, Spain} 
\affil[2]{Universidad de La Laguna, Spain}
\affil[3]{Universidad de Granada, Spain}
\affil[4]{Instituto Carlos I de F\'isica Te\'orica y Computacional, Spain}
\affil[5]{Università degli Studi di Padova, Italy}
\affil[6]{Nicolaus Copernicus Astronomical Center, Polish Academy of Sciences, Poland}
\affil[7]{Universidade Federal do Rio Grande do Sul (UFRGS), Brazil}
\affil[8]{Curtin Institute of Radio Astronomy, Australia}
\affil[9]{University of Vienna, Austria}
\affil[10]{Max-Planck-Institut f\"{u}r Astronomie, Germany}
\affil[11]{Instituto de Astrofísica de Andalucía, Spain} 
\affil[12]{Universidad Complutense de Madrid, Spain}
\affil[13]{Isaac Newton Group of Telescopes, Spain}

\begin{document}
\maketitle

\newpage
\begin{tcolorbox}[colback=RoyalBlue!5!white,colframe=black!75!black, width=\textwidth]
\justifying
{Nuclear star clusters (NSCs) are dense, compact stellar systems only a few parsecs across, located at galaxy centers. Their small sizes make them difficult to resolve spatially. NSCs often coexist with massive black holes, and both trace the dynamical state and evolution of their host galaxies. Dense stellar environments such as NSCs are also ideal sites for forming intermediate-mass black holes (IMBHs). 
To date, spatially resolved NSC properties, crucial for reconstructing dynamical and star-formation histories, have only been obtained for galaxies within 5~Mpc, using the highest-resolution instruments on the current class of very large telescopes. This severely limits spectroscopic studies, and a systematic, unbiased survey has never been accomplished. \textit{
Because the vast majority of known NSCs are located in the Northern Hemisphere, only a 30-m–class telescope in the North can provide the statistical power needed to study their physical properties and measure the mass of coexisting central black holes. \textbf{We propose leveraging the capabilities of a 30-m-class Northern telescope to obtain the first comprehensive, spatially resolved survey of NSCs, finally allowing us to unveil their formation pathways and their yet unknown connection with central massive black holes.}}}

\end{tcolorbox}

\section{Introduction and Motivation}
NSCs are among the densest stellar systems, only few parsecs across and with stellar masses from $10^4$ to several $10^8$\msun\,\citep[e.g.,][]{Hoyer2023a}. Located at galaxy centers, they often coexist with massive black holes (MBHs, $ \geq 10^3$\msun), and their stellar content traces the mechanisms that drive stars and gas into galactic nuclei \citep{Neumayer2020}. Two non-mutually exclusive formation channels are commonly invoked: star-cluster inspiral through dynamical friction \citep{Tremaine1975} and in situ star formation after gas inflow \citep[e.g.,][]{Milosavljevic2004}. 
Observations point to a combination of both \citep[e.g.,][]{Fahrion2024}. 

NSCs occur in more than 60\% of galaxies with stellar masses between $10^8$ and $10^{10}$\msun{}, peaking at 90\% near $10^{9}$\msun{} \citep{Hoyer2021}. Their frequency drops in low-mass systems and in high-mass early-type galaxies (ETGs), while remaining high in massive late types (LTGs) \citep{Neumayer2020}. These mass-dependent trends, together with chemical and age distributions \citep[e.g.,][]{Fahrion2022a}, suggest a transition from cluster-dominated at low masses to in situ formation at higher mass, closely tied to the presence and growth of their MBHs. 
Yet the coexistence (or absence) of MBHs in NSC-hosting galaxies remains unsolved, as well as the possibility that  intermediate-mass black holes (IMBHs) in low-mass galaxies are not absent but undetectable with current observatories. 
Kinematic and stellar population studies show younger, rotation-dominated NSCs in LTGs, mostly assembled via in situ star formation, while those in ETGs exhibit slower rotation and more complexity, resulting from past mergers \citep[e.g.,][]{Rossa2006,Pinna2021}. NSC incidence also increases in dense environments \citep{SanchezJanssen2019}, linking NSC formation to broader galaxy evolution. 
Another open question is whether NSCs, located at the galactic center, contain the first stars that formed in a galaxy, making them key to understanding the first stages of galaxy formation.

Theoretical studies reproduce aspects of both NSC formation pathways, but systematic simulations across galaxy mass, morphology, and environment remain scarce. 
High-resolution $N$-body models support the efficiency of cluster inspiral, while hydrodynamical simulations show that in situ star formation is required to reproduce observations \cite[e.g.,][]{Mastrobuono2014, Mastrobuono2019}. Although larger samples and a cosmological framework are needed to evaluate the balance of internal and external drivers of NSC and MBH growth, current cosmological simulations are still limited in resolution and in their ability to recover realistic galaxies. Thus, observations remain essential for constraining NSC formation pathways, yet detailed spectroscopic mapping is currently only feasible for very nearby systems (within $\sim$5 Mpc) using the most sensitive, large-aperture facilities \citep[e.g.,][]{Pinna2021}.

\section{The Science Challenge}
With current facilities, parsec-scale stellar mapping in NSCs is feasible only within $\sim$5 Mpc, which limits the number of currently classified NSCs to fewer than 25, and to 11 if we only consider the ones with parsec-scale integral-field spectroscopy (IFS) to date. Likewise, mass measurements of black holes (BHs), especially low-to-intermediate mass ones, require resolving their small spheres of influence, demanding angular resolutions of $\lesssim$0.02", thus limiting observations to galaxies closer than 10\,Mpc. A statistically robust picture of NSC formation and its link to galaxy evolution requires data for a larger, more diverse and unbiased galaxy sample, allowing for the investigation of trends with host-galaxy properties. 
Extending detailed NSC studies to a distance of 20\,Mpc 
will increase the number of reachable NSCs by a factor of more than 20 (visible either from the North or the South). 

About 70\% of the total sample of detected NSCs in these catalogs, hosted by galaxies of different masses, morphological types and environments (in the field, in galaxy groups, and in the Virgo Cluster), will not be visible from the Extremely Large Telescope (ELT) since they are located in the Northern Hemisphere. 
Therefore, a 30-m-class telescope in the North is essential for building this unbiased dataset. 
Of the very large sample visible from the North (from the latitude of the Roque de los Muchachos Observatory, in La Palma, but 
now including the ones with declinations above $-20^{\circ}$, visible also from the South), $\sim$60\% of NSCs are located within 20~Mpc. Such a facility will enable the first systematic and statistically significant survey of an unprecedentedly large and unbiased sample of 334 NSCs, with high-quality spatially resolved kinematics and stellar populations. 
The galaxy sample spans the full stellar-mass range of nucleated galaxies, between $\sim 10^6$ and $\sim 4\times 10^{11}$~\msun\, (with most galaxies around $10^{9}$~\msun, where the nucleated fraction is maximum), allowing to explore with IFS the lowest and highest mass ranges, which is mostly uncharted territory. 
Within our sample, approximately half are members of the Virgo cluster, and the rest are field or group galaxies. 
It will finally be possible to quantify the relative contributions of in situ star formation and cluster inspiral to NSC assembly \cite{Fahrion2022a} and to assess how these channels vary with galaxy mass, morphology, and environment. 
Determination of central BH masses across a wide galaxy parameter space will complement crucial insight into NSC–BH coevolution.

\subsection{Detailed zoom-in analysis of NSC}
\begin{figure}
\centering
\resizebox{0.99\textwidth}{!}
{\includegraphics[scale=1.]{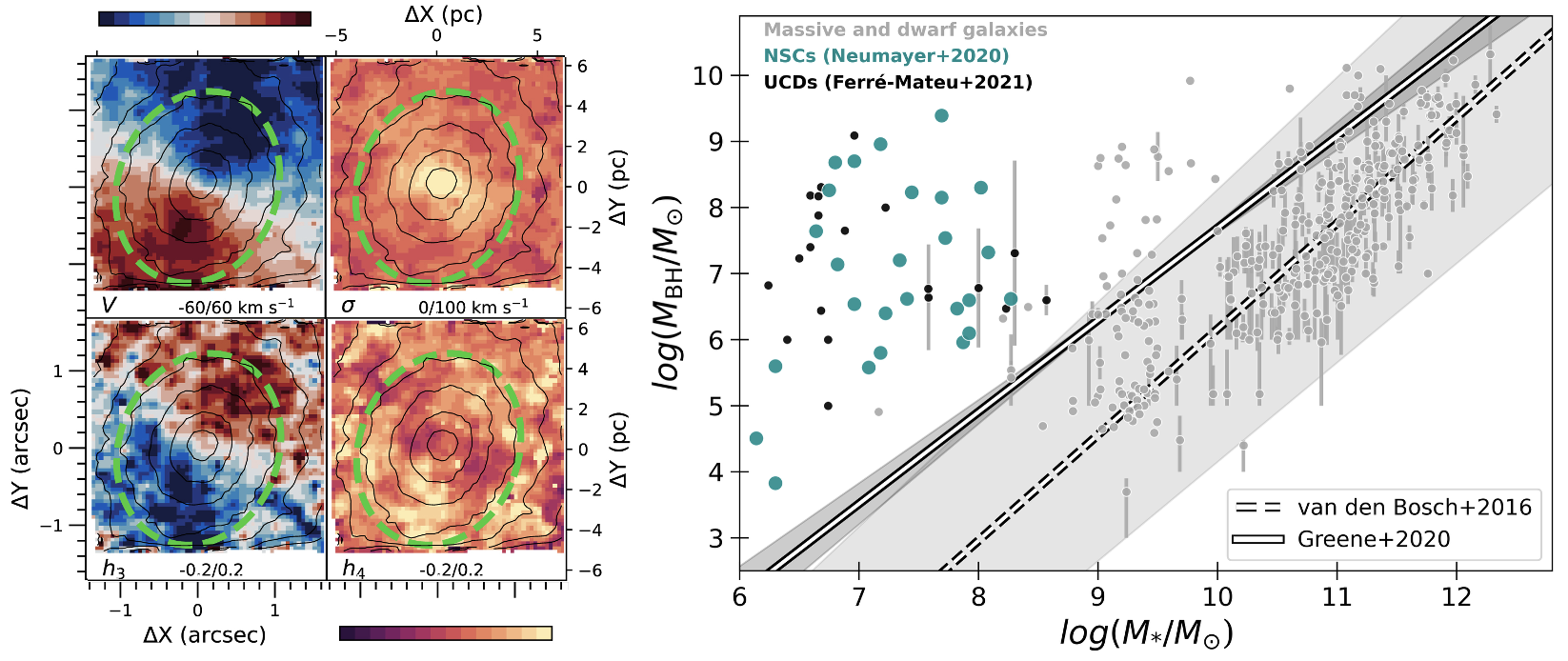}}
\caption{\label{figure}
\textit{Left panel:} Stellar kinematic maps of the NSC in M~32 adapted from \cite{Pinna2021}. From left to right, from top to bottom, mean velocity $V$, velocity dispersion $\sigma$, skewness $h_3$ and kurtosis $h_4$. 
Isophotes are shown in black and the NSC effective-radius ellipse in dashed green. 
\textit{Right panel:} BH mass versus stellar masses of NSCs and a variety of galaxy types, from the low mass ultracompact dwarfs, to the most massive systems, adapted from \cite{FerreMateu2021}.} 
\label{fig:image}
\end{figure}

With such a facility, we will be able to spatially resolve NSC kinematics and stellar populations in a large sample of galaxies (Fig.~\ref{figure}, left panel), allowing us to measure gradients within the NSC-dominated region and identify complex NSC structures with different components in the same NSC. 
Kinematically decoupled components in an NSC will trace the role of star or star cluster accretion, and thus galaxy mergers, in the formation of the first NSC seed and in its growth (see more examples for very nearby galaxies in \cite{Pinna2021}). 
Different stellar population components dominating age and metallicity maps at different radii will thus uncover substructures of various sizes within the same NSC, formed in different epochs of galaxy evolution. This has been done with current facilities for larger structures such as nuclear disks, unveiling the signatures of specific scenarios such as inside-out formation \citep[][]{Bittner2020}. The comparison with the surrounding galaxy will be key to linking with the host-galaxy evolution \cite{Pinna2025}. 
\subsection{Revealing hidden IMBHs and expanding direct BH measurements}
NSCs are the most promising environments to search for IMBHs and to constrain the low-mass end of the BH occupation fraction. 
Parsec-scale kinematic mapping within BH spheres of influence will enable robust dynamical mass measurements \citep{Nguyen2017, Nguyen2018, Nguyen2019} across a broad range of galaxy masses and morphologies, finally revealing BHs in the elusive IMBH regime. This capability, recently investigated for ELT/HARMONI \citep{Nguyen2025}, is essential because the low-mass end of MBH–galaxy scaling relations remains sparsely populated (see Fig.~\ref{figure}, right panel). The diversity of low-mass galaxy types produces significant scatter and/or apparent flattening below 
$\sim 10^{9}$\msun\,(e.g., \cite{FerreMateu2021}), with NSCs and other compact galaxies such as ultracompact dwarfs, playing a key role in driving such complexity. However, the gap seen at the lowest stellar masses might be mostly driven by the current telescope limitations for resolving and measuring the smallest MBHs. Accurate kinematics represent the most efficient method for finding IMBHs, as IMBHs appear to have exceptionally low accretion efficiencies compared to other classes of BHs, making them challenging to detect through multiwavelength observations \citep{Mahida2025}. 

\subsection{Capability Requirements}
NSCs have small sizes (between 1 and $\sim$20~pc, \cite{Hoyer2023a}) and low stellar velocity dispersion (down to $\lesssim10$~km s$^{-1}$, \cite{Pinna2021}). Their surface brightness can drop to $\gtrsim 22$~mag arcsec$^{-1}$ \citep{Carson2015} at the edges of the very central region where they dominate, and this makes it challenging to obtain the required signal-to-noise ratio for stellar-population mapping using high-resolution instruments (with small spaxels due to the high spatial sampling). 
Thus, spatially resolving NSC properties requires a specific set of instrumental capabilities beyond what 8–10 m telescopes can provide: 
\begin{itemize}
\item Milliarcsec angular resolution, at the diffraction limit of a $>$30-m telescope using adaptive optics (AO), to resolve sub-parsec spatial scales at distances of up to 20~Mpc. 
\item Advanced AO performance (with high Strehl ratios) and point-spread-function (PSF) stability, to preserve spatial resolution across the field of view. This is critical to fully exploit the very high spatial resolution and to avoid systematic errors in the derived properties.
\item Relatively high spectral resolution (R>8000) to measure 
NSC velocity dispersions of 10–30 km s$^{-1}$, $h_3$ and $h_4$, 
with the required accuracy for BH-mass measurements. This cannot be achieved from space, making a 30-40~m ground-based telescope the only way forward. 

\end{itemize}
\textbf{A 30-m-class telescope} represents the minimal infrastructure capable of meeting these requirements. It is therefore \textbf{the essential enabling technology for transforming NSC studies into a precision field and resolving the long-standing question of IMBH occupation in low-mass galaxies. 
This will significantly advance our understanding of black-hole demographics and formation, NSC assembly, and their co-evolution with host galaxies, thereby filling a critical gap in our understanding of galaxy evolution. }


\begin{multicols}{2}
\renewcommand{\bibfont}{\footnotesize}  
\bibliographystyle{unsrtnat}     
\bibliography{fp_nsc_short}

@ARTICLE{Mahida2025,
       author = {{Mahida} et al.},
      journal = {arXiv e-prints},
         year = 2025,
          eid = {arXiv:2512.09649},
        pages = {arXiv:2512.09649},
archivePrefix = {arXiv},
       eprint = {2512.09649},
 primaryClass = {astro-ph.HE},
       adsurl = {https://ui.adsabs.harvard.edu/abs/2025arXiv251209649M}
}

@ARTICLE{Bittner2020,
       author = {{Bittner} et al.},
      journal = {\aap},
         year = 2020,
       volume = {643},
          eid = {A65},
        pages = {A65},
archivePrefix = {arXiv},
       eprint = {2009.01856},
 primaryClass = {astro-ph.GA},
       adsurl = {https://ui.adsabs.harvard.edu/abs/2020A&A...643A..65B}
}

@ARTICLE{Carson2015,
       author = {{Carson} et al.},
      journal = {\aj},
         year = 2015,
       volume = {149},
       number = {5},
          eid = {170},
        pages = {170},
archivePrefix = {arXiv},
       eprint = {1501.05586},
 primaryClass = {astro-ph.GA},
       adsurl = {https://ui.adsabs.harvard.edu/abs/2015AJ....149..170C}
}

@ARTICLE{Fahrion2022a,
       author = {{Fahrion} et al.},
      journal = {\aap},
         year = 2022,
       volume = {658},
          eid = {A172},
        pages = {A172},
archivePrefix = {arXiv},
       eprint = {2112.05610},
 primaryClass = {astro-ph.GA},
       adsurl = {https://ui.adsabs.harvard.edu/abs/2022A&A...658A.172F}
}

@ARTICLE{Fahrion2024,
       author = {{Fahrion} et al.},
      journal = {\aap},
         year = 2024,
       volume = {687},
          eid = {A83},
        pages = {A83},
archivePrefix = {arXiv},
       eprint = {2404.08910},
 primaryClass = {astro-ph.GA},
       adsurl = {https://ui.adsabs.harvard.edu/abs/2024A&A...687A..83F}
}

@ARTICLE{FerreMateu2021,
       author = {{Ferr{\'e}-Mateu}, A. and {Mezcua}, M. and {Barrows}, R.~S.},
      journal = {\mnras},
         year = 2021,
       volume = {506},
       number = {4},
        pages = {4702-4714},
archivePrefix = {arXiv},
       eprint = {2107.02141},
 primaryClass = {astro-ph.GA},
       adsurl = {https://ui.adsabs.harvard.edu/abs/2021MNRAS.506.4702F}
}

@ARTICLE{Hoyer2021,
       author = {{Hoyer} et al.},
      journal = {\mnras},
         year = 2021,
       volume = {507},
       number = {3},
        pages = {3246-3266},
archivePrefix = {arXiv},
       eprint = {2107.05313},
 primaryClass = {astro-ph.GA},
       adsurl = {https://ui.adsabs.harvard.edu/abs/2021MNRAS.507.3246H}
}

@ARTICLE{Hoyer2023a,
       author = {{Hoyer} et al.},
      journal = {\mnras},
         year = 2023,
       volume = {520},
       number = {3},
        pages = {4664-4682},
archivePrefix = {arXiv},
       eprint = {2212.04151},
 primaryClass = {astro-ph.GA},
       adsurl = {https://ui.adsabs.harvard.edu/abs/2023MNRAS.520.4664H}
}

@ARTICLE{Mastrobuono2014,
       author = {{Mastrobuono-Battisti} et al.},
      journal = {\apj},
         year = 2014,
       volume = {796},
       number = {1},
          eid = {40},
        pages = {40},
archivePrefix = {arXiv},
       eprint = {1403.3094},
 primaryClass = {astro-ph.GA},
       adsurl = {https://ui.adsabs.harvard.edu/abs/2014ApJ...796...40M}
}

@ARTICLE{Mastrobuono2019,
       author = {{Mastrobuono-Battisti} et al.},
      journal = {\mnras},
         year = 2019,
       volume = {490},
       number = {4},
        pages = {5820-5831},
archivePrefix = {arXiv},
       eprint = {1910.11347},
 primaryClass = {astro-ph.GA},
       adsurl = {https://ui.adsabs.harvard.edu/abs/2019MNRAS.490.5820M}
}

@ARTICLE{Neumayer2020,
       author = {{Neumayer} et al.},
      journal = {\aapr},
         year = 2020,
       volume = {28},
       number = {1},
          eid = {4},
        pages = {4},
archivePrefix = {arXiv},
       eprint = {2001.03626},
 primaryClass = {astro-ph.GA},
       adsurl = {https://ui.adsabs.harvard.edu/abs/2020A&ARv..28....4N}
}

@ARTICLE{Milosavljevic2004,
       author = {{Milosavljevi{\'c}}, Milo{\v{s}}},
      journal = {\apjl},
         year = 2004,
       volume = {605},
       number = {1},
        pages = {L13-L16},
archivePrefix = {arXiv},
       eprint = {astro-ph/0310574},
 primaryClass = {astro-ph},
       adsurl = {https://ui.adsabs.harvard.edu/abs/2004ApJ...605L..13M}
}

@ARTICLE{Nguyen2017,
       author = {{Nguyen} et al.},
      journal = {\apj},
         year = 2017,
       volume = {836},
       number = {2},
          eid = {237},
        pages = {237},
archivePrefix = {arXiv},
       eprint = {1610.09385},
 primaryClass = {astro-ph.GA},
       adsurl = {https://ui.adsabs.harvard.edu/abs/2017ApJ...836..237N}
}

@article{Nguyen2018,
	year = 2018,
	publisher = {American Astronomical Society},
	volume = {858},
	number = {2},
	pages = {118},
	author = {Nguyen et al.},
	journal = {\apj}
}

@ARTICLE{Nguyen2019,
       author = {{Nguyen} et al.},
      journal = {\apj},
         year = 2019,
       volume = {872},
       number = {1},
          eid = {104},
        pages = {104},
archivePrefix = {arXiv},
       eprint = {1901.05496},
 primaryClass = {astro-ph.GA},
       adsurl = {https://ui.adsabs.harvard.edu/abs/2019ApJ...872..104N}
}

@ARTICLE{Nguyen2025,
       author = {{Nguyen} et al.},
      journal = {\aj},
         year = 2025,
       volume = {170},
       number = {2},
          eid = {124},
        pages = {124},
          doi = {10.3847/1538-3881/ade9ba},
archivePrefix = {arXiv},
       eprint = {2408.00239},
 primaryClass = {astro-ph.GA},
       adsurl = {https://ui.adsabs.harvard.edu/abs/2025AJ....170..124N}
}

@ARTICLE{Pinna2021,
       author = {{Pinna} et al.},
      journal = {\apj},
         year = 2021,
       volume = {921},
       number = {1},
          eid = {8},
        pages = {8},
archivePrefix = {arXiv},
       eprint = {2107.08903},
 primaryClass = {astro-ph.GA},
       adsurl = {https://ui.adsabs.harvard.edu/abs/2021ApJ...921....8P}
}

@ARTICLE{Pinna2025,
       author = {{Pinna} et al.},
      journal = {arXiv e-prints},
         year = 2025,
          eid = {arXiv:2512.03999},
        pages = {arXiv:2512.03999},
archivePrefix = {arXiv},
       eprint = {2512.03999},
 primaryClass = {astro-ph.GA},
       adsurl = {https://ui.adsabs.harvard.edu/abs/2025arXiv251203999P}
}

@ARTICLE{Rossa2006,
       author = {{Rossa} et al.},
      journal = {\aj},
         year = 2006,
       volume = {132},
       number = {3},
        pages = {1074-1099},
archivePrefix = {arXiv},
       eprint = {astro-ph/0604140},
 primaryClass = {astro-ph},
       adsurl = {https://ui.adsabs.harvard.edu/abs/2006AJ....132.1074R}
}

@ARTICLE{SanchezJanssen2019,
       author = {{S{\'a}nchez-Janssen} et al.},
      journal = {\apj},
         year = 2019,
       volume = {878},
       number = {1},
          eid = {18},
        pages = {18},
archivePrefix = {arXiv},
       eprint = {1812.01019},
 primaryClass = {astro-ph.GA},
       adsurl = {https://ui.adsabs.harvard.edu/abs/2019ApJ...878...18S}
}

@ARTICLE{Tremaine1975,
       author = {{Tremaine} et al.},
      journal = {\apj},
         year = 1975,
       volume = {196},
        pages = {407-411},
       adsurl = {https://ui.adsabs.harvard.edu/abs/1975ApJ...196..407T}
}
\end{multicols}

\end{document}